\documentclass{aa}
\usepackage[varg]{txfonts}
\usepackage[english]{babel}
\usepackage{amsmath}
\usepackage{graphicx}
\usepackage{hyperref}

\begin{document}

\title{Disk masses in the Orion Molecular Cloud-2: distinguishing time and environment}
\titlerunning{Disk masses in the Orion Molecular Cloud-2}

\author{S.E.~van Terwisga\inst{\ref{leiden1}}\and A. Hacar\inst{\ref{leiden1}}\and E.F. van Dishoeck\inst{\ref{leiden1},\ref{garching}}}

\institute{
Leiden Observatory, Leiden University, PO Box 9513, 2300 RA Leiden, The Netherlands\label{leiden1}\\ \email{terwisga@strw.leidenuniv.nl}\and
Max-Planck-Institut f{\"u}r Extraterrestrische Physik, Gie{\ss}enbachstraße, D-85741 Garching bei M{\"u}nchen, Germany\label{garching}
}

\abstract{The mass evolution of protoplanetary disks is driven by both internal processes, and external factors, such as photoevaporation. Disentangling these two effects, however, has remained difficult.}
{We measure the dust masses of a sample of 132 disks in the Orion Molecular Cloud (OMC)-2 region, and compare them to (i) externally photoevaporated disks in the Trapezium cluster, and (ii) disks in nearby low-mass star forming regions (SFRs). This allows us to test if initial disk properties are the same in high- and low-mass SFRs, and enables a direct measurement of the effect of external photoevaporation on disks.}
{A $\sim 20' \times 4'$ mosaic of 3\,mm continuum observations from the Atacama Large Millimeter/submillimeter Array (ALMA) was used to measure the fluxes of $132$ disks and $35$ protostars $>0.5$\,pc away from the Trapezium. We identify and characterize a sample of 34 point sources not included in the {\it Spitzer} catalog on which the sample is based.}
{Of the disks, 37 ($28\%$) are detected, with masses ranging from $7 - 270\,M_{\oplus}$. The detection rate for protostars is higher, at $69\%$. Disks near the Trapezium are found to be less massive by a factor $0.18^{+0.18}_{-0.11}$, implying a mass loss rate of $8 \times 10^{-8}\,M_{\odot}\,\text{yr}^{-1}$.}
{Our observations allow us to distinguish the impact of time and environment on disk evolution in a single SFR. The disk mass distribution in OMC-2 is statistically indistinguishable from that in nearby low-mass SFRs, like Lupus and Taurus. We conclude that age is the main factor determining the evolution of these disks. This result is robust with respect to assumptions of dust temperature, sample incompleteness and biases. The difference between the OMC-2 and Trapezium cluster samples is consistent with mass loss driven by far-ultraviolet radiation near the Trapezium. Together, this implies that in isolation, disk formation and evolution proceed similarly, regardless of cloud mass.}


\maketitle

\section{Introduction}
Protoplanetary disks are the environment in which planetary systems take shape: planets migrate, accrete mass in various parts of the disk, and in turn they can affect the disk itself, for instance by carving out gaps or rings. Throughout these processes, the total mass of the material in the disk is a key parameter: it sets the amount of material left to accrete into planetesimals; it also plays a role in determining the temperature profile of the disk (for a given surface density profile); likewise, the surface density at a given radius determines how rapidly a planet will migrate~\citep{benz14,baruteau14}. How the masses of protoplanetary disks evolve, and are affected by environmental factors, is therefore a fundamental question to answer for understanding planet origins.

In recent years, the Atacama Large Millimeter/submillimeter Array (ALMA) and Submillimeter Array (SMA) telescopes,  with their excellent sensitivities, have been essential in revealing the mass distributions of cold dust in large samples of protoplanetary disks in nearby star-forming regions (SFRs). Thanks to the presence of very complete and deep censuses of disk host stars, made primarily with {\it Spitzer}, these surveys are also unbiased, reaching warm dust masses as low as $0.1$ lunar mass~\citep{evans03}. This completeness makes it possible to compare the disk mass distribution across regions.

Surveys to date span a wide range of physical conditions and ages, with the $\sigma$ Ori~\citep{ansdell17} and Orion Nebula Cluster or Trapezium cluster~\citep[e.g.,][]{eisner08,mann10} (ONC) typical of massive, young SFRs. Regions like Lupus, Taurus and Chamaeleon I on the other hand probe a lower-mass, more isolated regime of star formation, at ages between $1-5$\,Myr~\citep{ansdell16,andrews13,pascucci16}. Observations of the Upper Scorpius OB association (Upper Sco), meanwhile, provide a window on a more evolved population~\citep{barenfeld16}.

From the comparison of the disk masses presented in these surveys, two results become apparent. First, in low-mass star-forming regions where stars and their disks form and evolve in relative isolation, the disk mass distributions of regions of similar ages are essentially indistinguishable. For example, Taurus and Lupus disks are statistically the same in terms of dust mass distribution~\citep{ansdell16}. With age, a steep decline in the median disk mass is seen, by a factor $\sim 3$, although some massive disks persist up to 10\,Myr in Upper Sco~\citep{barenfeld16}.

However, in the dense environment of the ONC and in $\sigma$ Ori, the presence of young massive stars can have a dramatic effect: even a young population ($\sim 1-3$\,Myr) like the ONC has disks that are significantly less massive than their counterparts in Lupus~\citep{eisner18}. Also, disk mass decreases in the vicinity of the Trapezium cluster~\citep{mann14} and $\sigma$ Ori~\citep{ansdell17}. These results, as well as observations of proplyd structures in the inner ONC~\citep[e.g.,][]{odell93,odell98}, have led to the conclusion that externally-driven photoevaporation is the main cause for the extra mass loss in these objects.

The purpose of the survey presented in this article is to characterize the bulk dust mass contained in cold, mm-sized grains in a population of disks around young stars centered on the Orion Molecular Cloud (OMC) 2 region in the Orion A cloud, away from the strong UV radiation near the Trapezium~\citep{peterson08}. This is done using observations from the Atacama Large Millimeter/submillimeter Array (ALMA). 

Environments like the Orion nebula play an important role in our understanding of planet formation because they are such common sites of star formation, with up to half the stars in the Galaxy forming in similar high-mass SFRs~\citep[e.g.,]{lada93, carpenter00}, likely including the Sun~\citep{williams07}. Although the disk population in the densest parts of Orion is now well-studied, the properties of disks in the \emph{outer} regions are in comparison much less well-constrained. For instance, in~\citet{mann10} and~\citet{mann14}, disks out to 2\,pc from the Trapezium are sampled, and reveal a clear decrease of disk mass with proximity to the ONC; but the population beyond 0.5\,pc is sampled with fewer than 10 disks.

\citet{megeath12} used {\it Spitzer} observations to show a significant population of disks with warm dust extending far beyond the ONC, and running along the Integral-Shaped Filament (ISF). The disks in OMC-2 offer a unique view of the properties of disks in a massive star-forming region: here, bright young O-type stars are absent, leading to a much lower interstellar radiation field, which approaches the conditions in low-mass star-forming regions. This allows us to constrain, for the first time, a population of disks unaffected by external photoevaporation but still in relatively close proximity to the Trapezium cluster. By comparing disk properties between this sample and, e.g., the Lupus population, it is possible to gain insight in the initial conditions of disk formation in a massive cloud.

The large mosaic of ALMA 12-m data of the OMC-1 and OMC-2 regions that form the northern part of the ISF, presented in~\citet{hacar18}, is a uniquely valuable resource in this context. Inside 0.5\,pc from the Trapezium, heavy contamination from partially resolved-out continuum and free-free emission is present. However, these observations cover a large area containing the population of interest for this study, which is located beyond 0.5\,pc from the Trapezium. {\it Spitzer} data of this region allow us to extract disk properties in a homogeneous and unbiased manner.

In this article, we present the first large ($N=132$) blind unbiased survey of disks in OMC-2, centered on the ISF, and use this sample to infer the rate of mass loss in the vicinity of the Trapezium in a consistent manner. We also compare the properties of this sample to that in nearby star-forming regions of similar ages in order to test if the initial conditions for disk formation are the same everywhere. We find that disk masses in this environment strongly resemble those in the nearby, low-mass Lupus and Taurus star-forming regions, but not those in the ONC, suggesting that disk formation and evolution may proceed similarly regardless of the mass of the cloud from which they form.

\section{Observations and reference catalog}
\subsection{ALMA data}
The data at the basis of this project are 3\,mm ALMA Band-3 continuum observations that were taken as part of observing program 2015.1.00669.S (PI: A. Hacar). The primary purpose of this dataset was to collect information on the large-scale distribution of N$_{2}$H$^{+}\,(1-0)$ at 93.2\,GHz in Orion; these results are described in~\citet{hacar18}. Throughout this paper, we will be using the same distance to the OMCs, of 414\,pc~\citep{menten07}. Our data analysis relies on the same calibrators (J0423-0120 for amplitude and bandpass; J0541-0541 for phase), and covers the same area in a mosaic: a field of $\sim 20'$ in declination and $\sim 4'$ in RA, spanning the ISF between the Trapezium and the southern part of the OMC-3 region, and centered on the OMC-2 cloud. The full image consists of two, 148-pointing sub-mosaics: one centered on the OMC-1 cloud and extending over the Trapezium and southern OMC-2, one on the OMC-2 and southern OMC-3 clouds. Of these sub-mosaics, the Southern fields generally have a lower noise level, due to a slightly deeper integration; the resulting difference in effective primary beam coverage in the full image is on the order of $\sim 30\%$ between the northern and southern sub-fields, which we take into account throughout the following by weighting the image noise with the primary beam coverage at all points.

The continuum data span three windows, of $1.8$\,GHz each, at 93, 104 and 108\,GHz. Some line contamination is present, and the corresponding spectral ranges are removed throughout the image. The rest of the data are combined to maximize the $S/N$ of our output images, leading to an effective frequency of 99.5\,GHz for the observations discussed here.\footnote{The reduced data for this project are available online, at the ORION-4D project website \url{https://sites.google.com/site/orion4dproject/home}.}

\subsection{A point-source map of OMC-2}
The original purpose of the survey data used here was to detect the extended emission of N$_{2}$H${^+}\,(1-0)$ across the cloud with a compact array configuration. This was achieved by using baselines between 3.5 and 92\,k$\lambda$. As a result, emission can be recovered up to scales of $\sim 1'$ from the 12-meter data alone. For this reason imaging and analysing the continuum from the ALMA 12-m array data products directly is not suitable for our purposes: contamination from partly resolved-out cloud material is a significant concern, especially for those objects that are more embedded.

\begin{figure*}[hp]
\begin{center}
		\includegraphics[width=\textwidth]{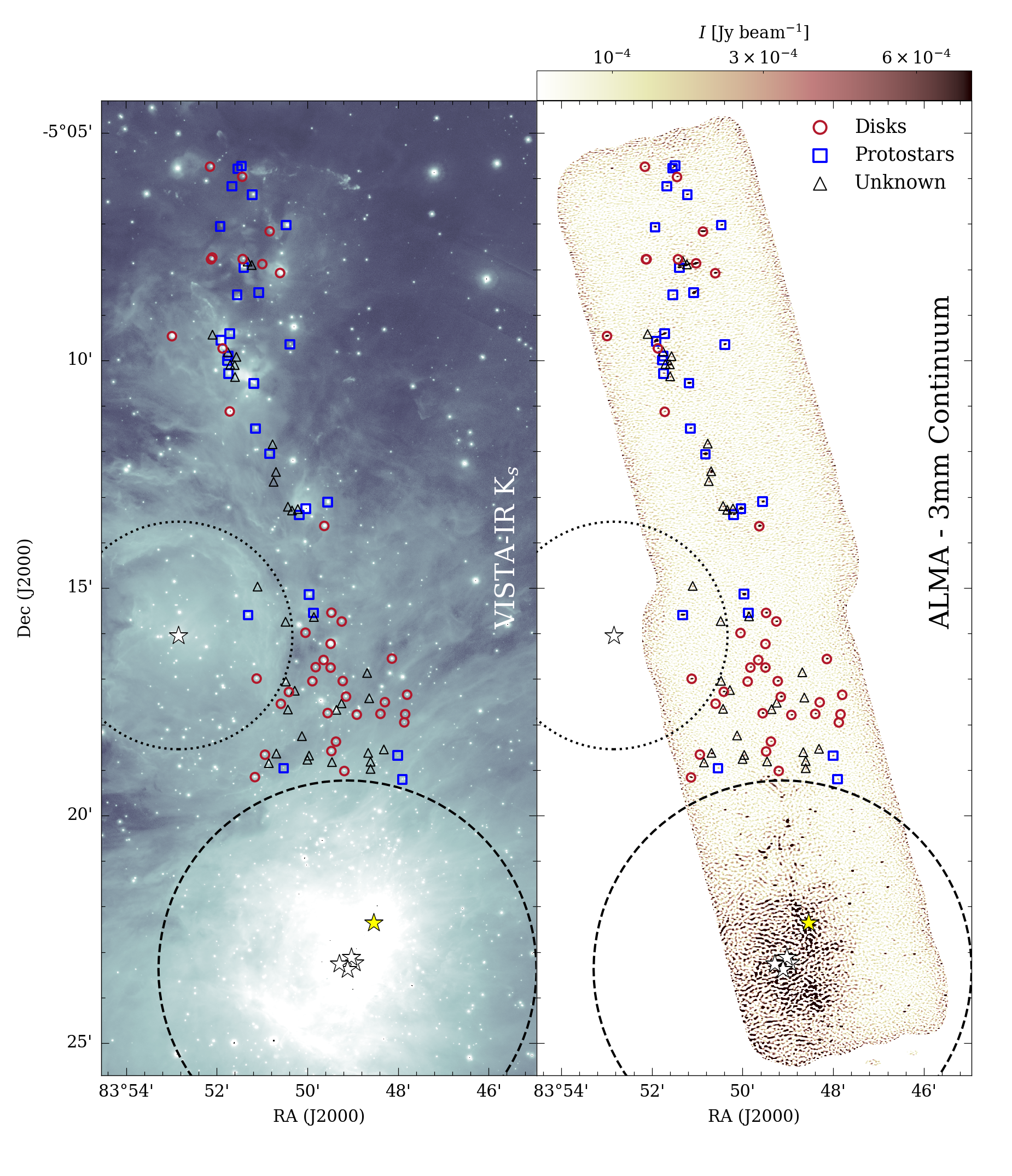}
		\caption{Spatially filtered ALMA observations of the OMC-2-region at 3\,mm {\it (right)} and VISTA K$_s$ observations of the same area {\it (left)}~\citep{meingast16}. Point sources detected in the ALMA data are marked: red circles and blue squares indicate disks and protostars from the~\citet{megeath12}-catalog respectively; sources without detected counterparts in this catalog are marked with black triangles. White stars indicate the locations of the Trapezium stars $\theta^1$ Ori A -- D, and HD 37061; the dashed black line indicates the photoionization region around the Trapezium out to 0.5\,pc, while M42 is marked by the dotted circle. The Orion KL outflow is shown by the yellow star.}
		\label{fig:bigmap}
\end{center}
\end{figure*}

Fortunately, the interferometric nature of the dataset allows us to efficiently reconstruct the data so that only mostly unresolved point sources are left in the resulting image, as done in other studies~\citep[e.g.,][]{felli93,eisner08,eisner18}. This is done by making a cut in the $u,v$-plane, and removing all baselines shorter than $30$\,k$\lambda$ prior to imaging, which leads to a largest resolvable scale of around $7''$ or $\sim 3000$\,AU at the $414$\,pc distance of Orion. This scale ensures that even the largest Class II disks are unresolved, while removing the extended cloud emission, and keeping good baseline coverage in $u,v$-space. 

The results of the imaging procedure are shown in Fig.~\ref{fig:bigmap}. On the left panel, the VISTA $K_s$-band image of the same region from~\citet{meingast16} is shown; the spatially-filtered ALMA map on the right shows the point sources in our field, leaving only a few (very young) sources marginally resolved. The absence of large-scale emission from this image is immediately obvious; however, numerous point sources are directly correlated to infrared (IR) detections of young stars in the VISTA data.

In order to image the data we used the CASA version 5.4.0 implementation of the CLEAN algorithm in the \texttt{tclean} task, to avoid a recently-discovered issue with flux retrieval for rectangular mosaics like this one in earlier CASA versions. This task's \texttt{mosaic} gridder option leads to the image used in this article, which has a $2.7'' \times 1.6''$ beam, and an effective $1\sigma$ noise level of $0.1$\,mJy\,beam$^{-1}$ in the southern (deeper) part of the image. Dust mass sensitivities are lower than in previous surveys, both due to the intrinsically fainter dust emission at longer wavelengths and the depth of these observations. We can detect sources at $3\sigma$ down to a dust mass of $17\,M_{\oplus}$~(see Sect.~\ref{sec:mdisk}), compared to $0.2\,M_{\oplus}$ in Lupus~\citep{ansdell16}. In the least sensitive parts of the northern sub-mosaic, sources down to $32\,M_{\oplus}$ can be detected. This dataset therefore allows us to sample the upper part of the disk mass distribution function, given a sufficiently large sample.

Around the ONC, significant extended emission and structures at all scales are present, and free-free contamination is important at the wavelenghts of our observations~\citep{dicker09}. We therefore avoid the primary photoionization region, which extends out to 0.5\,pc away from the Trapezium cluster. Despite the lower data fidelity we note that even in the Trapezium several point sources with clear IR counterparts can be identified by eye, but we do not analyze this region further, since, as Figure~\ref{fig:bigmap} shows, this area is strongly confused.

\subsection{{\it Spitzer} catalog of sources}
The purpose of this study is to compare the disk population in the OMC-2 region with that in other star-forming regions in a homogeneous manner. For this an unbiased and uncontaminated input catalog of YSOs is necessary. Moreover, the classification criteria of sources should be similar to those used in other studies of protoplanetary disks, which places strict limits on the possible base catalogs.

In this article, the base catalog is the {\it Spitzer} survey of young stellar objects in the Orion A and B clouds by~\citet{megeath12}. This catalog classifies young stellar objects based on their mid-infrared excess in a homogeneous way, and classifies them as Disks, Protostars, and Red and Faint Candidate Protostars respectively, with Disks being Class II YSOs and the other objects younger, or ambiguous in nature. Two objects are classified as Red Protostars in the field covered by our observations. These are included in the Protostars for the purposes of this survey. For some sources (see~\citealt{furlan16,kainulainen17}) the classification was changed based on subsequent analysis of the SED; we used the latest published source class in those cases, and updated the source identifications accordingly. Sources identified as protostars with a flat spectrum were not changed, since their nature is ambiguous. In all cases, this approach lead to the reclassification of a suspected disk source to a protostellar source; all reclassified sources (see below) were among the brightest detections, indicating that this is a conservative approach.

The~\citet{megeath12} {\it Spitzer} catalog of Orion YSOs is a powerful tool for the type of analysis used in this paper: it is consistent with the way other samples of Class II disks in nearby star-forming regions have been defined, and since the infrared emission from warm dust detected by {\it Spitzer} is optically thick it is able to detect disks even when disk masses are low; likewise, in theory, it is possible to detect young stellar objects down to the hydrogen-burning limit at the distance of Orion. However, this catalog is not complete: it is limited both by the resolution of the telescope ($2'' - 5''$) and the completeness is sensitive to the nebulosity and stellar density in the area~\citep{megeath16,grossschedl18}, both factors which are not constant across the image.

\section{Results}
\subsection{A 3-mm catalog of YSOs in OMC-2}
The full ALMA continuum map of the OMC-2 region, shown in Figure~\ref{fig:bigmap}, combined with the {\it Spitzer} catalog, shows a number of clear detections. In agreement with~\citet{kainulainen17}, the protostars primarily cluster along the densest part of the ISF; these objects are also those most likely to show evidence of somewhat resolved structure (see for instance, in Fig.~\ref{fig:protstamps}, P1, and P10). The Class II disks appear to be more scattered over the field. To the south of the ISF, below $-5^{\circ} 15' 51''$, the number of protostars drops rapidly due to the lack of dense pre-stellar gas~\citep{hacar18}. In this region, there is little mm-wavelength contamination even prior to the spatial filtering used here, and the number of disks (in the catalog and in the mm-image) seems to go up, as does the number of sources that do not have a counterpart in the {\it Spitzer} catalog.

\subsubsection{{\it Spitzer}-detected sources}
Over the full field (excluding the ONC), we have sample sizes of $N=132$ for disks and $N=35$ for younger objects detected by {\it Spitzer}. Of the disks, 37 are detected by our ALMA observations at at least $3\sigma$, for a detection rate of $28\%$. Zoomed-in, $36'' \times 36''$ cutouts of all detected disks and protostars with {\it Spitzer} counterparts are shown in Appendix~\ref{app:cutouts}, as Figures~\ref{fig:diskstamps} and~\ref{fig:protstamps}. These have been ordered by flux for easier inspection; their disk- and protostellar fluxes can be found in Tables~\ref{tab:D} and~\ref{tab:P}, respectively. The full tables (including upper limits) are available on the CDS. One of the disks (D31) is a binary object with a recently reported variability at submillimeter wavelengths, and its mas should therefore be interpreted carefully~\citep{mairs19}.

The detection rate for protostars is significantly higher than for disks: 26 detections give a detection rate of $69\%$ for these objects. Both populations overlap in flux, but the median flux for protostars is higher than for disks, consistent with the loss of most of the envelope after Class I~\citep{tychoniec18}. Likewise, the protostars are more likely to be associated with resolved millimeter emission (Fig.~\ref{fig:protstamps}) than the disks in the sample (Fig.~\ref{fig:diskstamps}). Additional data are needed to constrain the mass properties of this protostellar population, preventing their comparison with our well-defined disk sample.

Comparing the results of this survey to those presented in~\citet{kainulainen17} shows that all protostars detected in their data as well as in the {\it Spitzer} catalog are also detected in our data, although the beam in the ALMA data analyzed by~\citet{kainulainen17} is somewhat larger ($3.75'' \times 2.27''$) and their $S/N$ is lower (at 0.23\,mJy\,beam$^{-1}$). Importantly, the field observed here is also a factor of 2 larger. While the overlap in protostellar sources is large, the data presented in this paper do not allow us to recover a number of starless cores. This is in good agreement with the expectation that such cores lack the compact structures we are looking for, and that they are filtered out by the imaging procedure used here.

Fluxes for all sources were measured by searching for the brightest pixel within a $2.5''$ radius of the catalog positions in the ALMA point-source map, to account for the relative beam sizes of both ALMA and Spitzer; noise levels were determined from the image noise weighted with the primary beam coverage at each position. Sources were considered detections in case their flux exceeded the $3\sigma$ noise. In order to ensure the detections are reliable, a simple visual check was also performed on each position, but no detections were removed or added to the sample as a result.

The brightest, and presumably most massive, sources in this sample are worth a closer look. For the disks, it is remarkable that the two brightest sources (D1 and D2, top rows of Fig.~\ref{fig:diskstamps}) at millimeter wavelengths are associated with very extincted regions of the VISTA $K_s$-image, and may be somewhat contaminated by larger-scale emission. However, in general, the 10 brightest disks are compact (indicating that they are not misclassified), and scattered throughout the field. These objects, then, are the equivalent of the most luminous mm-disks in nearby low-mass SFRs.

In the protostellar sample, it is apparent that the brightest protostars are located in the dense clumps of the ISF: these objects are surrounded by other compact sources at mm-wavelengths, a significant fraction of which are other protostars. These clumps are concentrated to the north of the surveyed area. There is, however, no statistically significant difference in luminosities of the northern population of protostars (corresponding roughly to the spine of the OMC-2 cloud, beyond 1\,pc from the Trapezium) and the protostars in the southern part of the field, given this limited sample size.

\begin{table*}
\caption{Continuum fluxes and masses for the detected disks}
\label{tab:D}
\centering
\begin{tabular}{l l l l l}
\hline \hline
ID & RA & Dec & Flux & Mass \\
 & J2000 & J2000 & mJy & $M_{\oplus}$ \\
\hline
D1 & 05:35:23.98 &-05:07:53.0 & $5.0 \pm 0.17$ & $273.1 \pm 9.78$ \\
D2 & 05:35:23.33 &-05:07:09.8 & $4.8 \pm 0.18$ & $259.9 \pm 10.30$ \\
D3 & 05:35:18.51 &-05:13:38.2 & $4.4 \pm 0.18$ & $240.4 \pm 10.42$ \\
D4 & 05:35:31.95 &-05:09:28.1 & $3.9 \pm 0.32$ & $209.5 \pm 18.21$ \\
D5 & 05:35:22.41 &-05:08:04.8 & $2.8 \pm 0.18$ & $146.6 \pm 10.22$ \\
D6 & 05:35:17.95 &-05:16:45.0 & $2.5 \pm 0.11$ & $127.2 \pm 6.59$ \\
D7 & 05:35:16.59 &-05:17:23.2 & $1.9 \pm 0.11$ & $97.5 \pm 6.32$ \\
D8 & 05:35:24.48 &-05:16:59.4 & $1.8 \pm 0.12$ & $91.9 \pm 6.80$ \\
D9 & 05:35:17.88 &-05:15:32.6 & $1.7 \pm 0.11$ & $81.2 \pm 6.12$ \\
D10 & 05:35:16.88 &-05:17:02.7 & $1.5 \pm 0.11$ & $74.3 \pm 6.36$ \\
D11 & 05:35:21.63 &-05:17:17.1 & $1.5 \pm 0.11$ & $72.9 \pm 6.41$ \\
D12 & 05:35:18.21 &-05:17:44.9 & $1.4 \pm 0.11$ & $67.1 \pm 6.52$ \\
D13 & 05:35:12.54 &-05:16:33.1 & $1.4 \pm 0.13$ & $64.5 \pm 7.68$ \\
D14 & 05:35:16.97 &-05:15:44.2 & $1.3 \pm 0.11$ & $59.9 \pm 6.34$ \\
D15 & 05:35:24.63 &-05:19:09.3 & $1.3 \pm 0.37$ & $58.2 \pm 21.32$ \\
D16 & 05:35:15.64 &-05:17:47.0 & $1.1 \pm 0.11$ & $49.8 \pm 6.39$ \\
D17 & 05:35:25.75 &-05:05:57.9 & $1.0 \pm 0.20$ & $44.5 \pm 11.44$ \\
D18 & 05:35:28.59 &-05:05:44.6 & $1.0 \pm 0.27$ & $44.4 \pm 15.24$ \\
D19 & 05:35:25.71 &-05:07:46.4 & $1.0 \pm 0.17$ & $42.9 \pm 10.01$ \\
D20 & 05:35:19.27 &-05:16:44.4 & $1.0 \pm 0.12$ & $40.5 \pm 6.64$ \\
D21 & 05:35:16.74 &-05:19:01.3 & $0.9 \pm 0.11$ & $40.0 \pm 6.29$ \\
D22 & 05:35:13.55 &-05:17:45.9 & $0.9 \pm 0.12$ & $39.8 \pm 6.67$ \\
D23 & 05:35:27.48 &-05:09:44.2 & $0.9 \pm 0.16$ & $37.0 \pm 9.41$ \\
D24 & 05:35:23.74 &-05:18:39.8 & $0.7 \pm 0.15$ & $28.4 \pm 8.40$ \\
D25 & 05:35:28.50 &-05:07:46.7 & $0.7 \pm 0.16$ & $28.3 \pm 9.22$ \\
D26 & 05:35:17.48 &-05:18:22.6 & $0.6 \pm 0.11$ & $22.4 \pm 6.41$ \\
D27 & 05:35:20.17 &-05:15:59.0 & $0.6 \pm 0.11$ & $21.2 \pm 6.32$ \\
D28 & 05:35:18.57 &-05:16:34.9 & $0.6 \pm 0.12$ & $18.6 \pm 6.70$ \\
D29 & 05:35:11.20 &-05:17:20.7 & $0.5 \pm 0.17$ & $16.4 \pm 9.65$ \\
D30 & 05:35:26.85 &-05:11:07.3 & $0.5 \pm 0.15$ & $13.9 \pm 8.74$ \\
D31\tablefootmark{a} & 05:35:17.95 &-05:16:13.6 & $0.5 \pm 0.12$ & $12.9 \pm 6.68$ \\
D32 & 05:35:11.45 &-05:17:57.1 & $0.5 \pm 0.13$ & $12.9 \pm 7.32$ \\
D33 & 05:35:19.56 &-05:17:02.9 & $0.5 \pm 0.11$ & $12.2 \pm 6.54$ \\
D34 & 05:35:17.88 &-05:18:35.0 & $0.4 \pm 0.11$ & $11.6 \pm 6.42$ \\
D35 & 05:35:11.37 &-05:17:46.3 & $0.4 \pm 0.14$ & $10.0 \pm 7.85$ \\
D36 & 05:35:22.34 &-05:17:32.7 & $0.4 \pm 0.11$ & $9.0 \pm 6.37$ \\
D37 & 05:35:13.16 &-05:17:30.5 & $0.4 \pm 0.12$ & $6.9 \pm 6.75$ \\
\hline
\end{tabular}
\tablefoot{
\tablefoottext{a}{Variable source at mm-wavelengths~\citep{mairs19}.}}
\end{table*}

\subsubsection{Non-{\it Spitzer} point sources}
\label{sec:nospitzer}
The effect of the limitations of the {\it Spitzer}-data is that we would underestimate the fraction of lower-mass disks in the sample, especially toward the south of the area covered: this region has the brightest background, and while more massive and luminous YSOs are still easily detectable, less-massive stars (with similarly fainter disks) can be absent from the sample by failing one or more inclusion criteria for the {\it Spitzer} catalog. In addition, the youngest (Class 0) objects are expected to be completely absent from the {\it Spitzer} data, as they are intrinsically very faint at infrared wavelengths.

As a result, a search for point sources not corresponding to any {\it Spitzer}-defined YSOs was carried out. The \texttt{peak\char`_local\char`_max}-function from \texttt{scikit-image} was used to identify point sources in the image but not in the {\it Spitzer} catalog, and fluxes were extracted using the same $2.5''$ aperture around the identified source position that was also used for the Disk- and Protostar samples. This sample -- the Unidentified point sources -- consists of $N=34$ sources. Like for the other objects, cutouts superimposing ALMA data over the VISTA K$_s$ observations from~\citet{meingast16} are included in Appendix~\ref{app:cutouts}; fluxes and (tentative) masses are listed in Table~\ref{tab:U}.

Classifying this sample is difficult: other surveys of YSOs in Orion do not cover the full area of our observations, or lack the depth to detect faint objects, leading to ambiguities. Here, the ALMA observations by~\citet{kainulainen17}, as well as the {\it Herschel} Orion Protostar Survey~\citep{furlan16}, the {\it Chandra} Orion Ultradeep Point Source Catalog~\citep{getman05} and the Vienna Survey In Orion (VISION)~\citep{meingast16} are used to provide context to the objects not included in the {\it Spitzer} catalog. We will implicitly assume that sources without counterparts in any of these surveys are still part of the Orion SFR: the number density of background galaxies is low, at $6.1 \pm 3.1$ per square degree or $< 0.023$ in our field of view~\citep{megeath12}.

By examining the distribution of the Unidentified sources over the cloud, it becomes apparent that this subsample seems to consist of two distinct parts. In the northern part of the image, detections are located along the ISF, in a similar manner to the known distribution of protostars along this structure. If these sources do correspond to young, very extincted sources, it is to be expected that this sample overlaps significantly with the youngest YSOs in~\citet{furlan16}, and this is indeed the case. While the overlap is not complete, a large number (10 of 17) of sources located north of $-5^{\circ} 15' 51''$ are also detected by~\citet{kainulainen17} and correspond to a PACS Bright Red Source (PBRS), Class I or Class 0, or starless cores. In contrast, most millimeter point sources in this area have no counterpart in the VISION survey data (see Fig.~\ref{fig:ustamps}). In particular, none of the Unidentified sources with counterparts in~\citet{kainulainen17} are detected in the VISION survey data, indicating their embedded nature. Thus, we do not consider these sources further in our analysis.

In contrast, toward the southern field, the distribution of Unidentified sources more closely resembles that of the Disks in that area. Not only are these objects more scattered over the field than their counterparts to the north, they seem to be fainter on average, although this may be an effect of the larger number of sources in the outskirts of the ONC. In this area of the map, the main sources of information are the COUPS~\citep{getman05} and VISION~\citep{meingast16} surveys. Both of these surveys -- at X-ray-wavelengths, and in the $K_s$-band, respectively -- are more suited to identifying Class II and Class III protostars. In line with the expectations for a subsample consisting mostly of previously unidentified disks, there is a large fraction of X-ray and near-infrared counterparts for these objects: within a $2.5''$ radius, $11/18$ sources have VISION counterparts, and $7/18$ have counterparts in the COUP survey. With a median age of 0.8\,Myr~\citep{getman05}, this suggests that Class II sources dominate the southern sample of otherwise unidentified ALMA point sources. In Sect.~\ref{sec:completeness}, the impact of adding these sources to the disk sample is quantified.

\subsection{Disk masses in the OMC-2 field}
\label{sec:mdisk}
In order to estimate the disk mass distribution, we must relate our observed 3\,mm fluxes to the disk mass, and to do so in a manner that is consistent with the literature samples this dataset will be compared to. Typically, the following equation is used~\citep[e.g.,][]{ansdell16, barenfeld16, eisner18}:
\begin{equation}
M_{\text{dust}} = \frac{d^2 F_{\nu,\,\text{dust}}}{\kappa_{\nu} B_{\nu}(T_{\text{eff}})},
\label{eq:thindisk}
\end{equation}
where the millimeter flux $F_{\nu}$ (corrected for free-free emission; see Sect.~\ref{sec:free-free}) is assumed to be due to blackbody emission from optically thin dust with an average temperature $T_{\text{eff}}$, and a known effective opacity $\kappa_{\nu} = \kappa_0 (\nu/\nu_0)^{\beta}$. For the dust opacity, usually, $\beta = 1$ is used and $\kappa_{1000\,\text{GHz}} = 10$\,cm$^{2}$\,g$^{-1}$~\citep{beckwith90}. The effective temperature of the emitting material is taken to be $T_{\text{eff}} = 20$\,K, even though one could argue a somewhat higher temperature ($T_{\text{eff}} = 30$\,K) (see Sect.~\ref{sec:tempeff}) is more plausible in a massive star-forming region like Orion. These are the same values that are used in the articles that present the disk masses to which we compare the OMC-2 sample, thus ensuring a consistent comparison of the samples.

\begin{figure}[ht]
\begin{center}
		\includegraphics[width=0.49\textwidth]{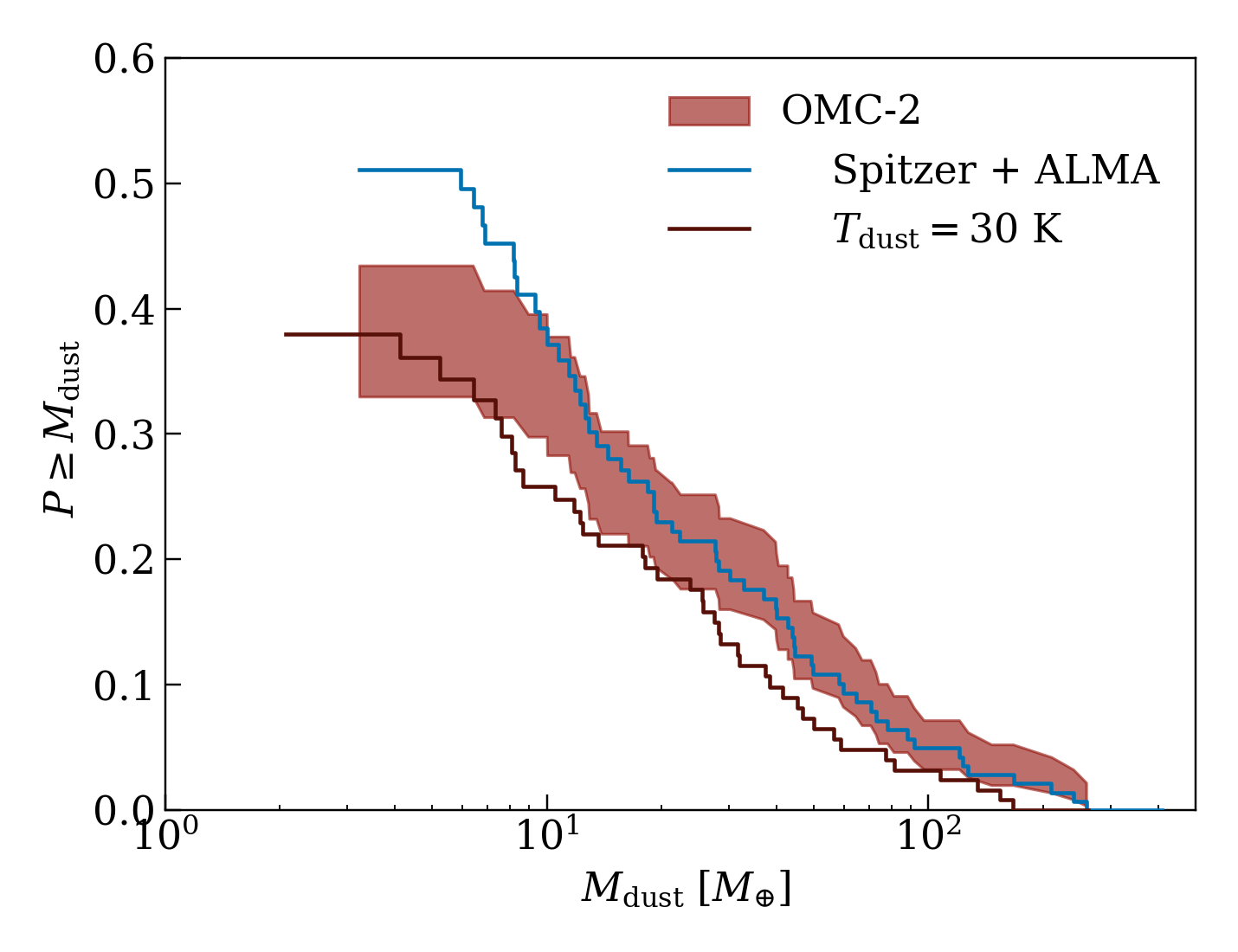}
		\caption{Inferred disk mass distribution for the OMC-2 disk population. The effect of adding the ALMA-detected point sources likely to be disks but missing in the {\it Spitzer} catalog is shown in blue; the red-brown line indicates the impact of a higher effective dust temperature of $30$\,K.}
		\label{fig:impact}
\end{center}
\end{figure}

Using a Kaplan-Meier estimator, it is then possible to estimate the disk mass distribution in the OMC-2 field while taking upper limits on the disk mass into account. The resulting distribution and its uncertainties are shown in Figure~\ref{fig:impact}. Despite the low fraction of detected sources of $29\%$, the uncertainty level is still excellent thanks to the large sample size. The disks observed clearly cover a large range of disk masses, between $273 \pm 9.8\,M_{\oplus}$ and $6.9 \pm 6.4\,M_{\oplus}$ (note that, for this latter value, the free-free contribution has been subtracted; see Sect.~\ref{sec:free-free}). The most massive $5\%$ of disks (6 sources) have masses in excess of $100\,M_{\oplus}$. In the following Sections, the robustness of this result is considered.

\subsection{Robustness of results}
\subsubsection{The effect of sample incompleteness}
\label{sec:completeness}
The Unidentified sources present in our ALMA map may be the result of the underlying biases in the {\it Spitzer} catalog~\citep{megeath16, grossschedl18}. While it is not possible to find the number of Class II disks not included in the {\it Spitzer} sample that are also undetected in these ALMA observations, the Unidentified sources allow us to constrain the impact of any biases, and to describe how this affects the comparison of this sample to that in other regions. As discussed in Sect.~\ref{sec:nospitzer}, most of the northern sources can be matched to Class I and Class 0 YSOs, but it is likely that the southern sources (below $-5^{\circ} 15' 51''$) are disks. If all of these sources are added to the {\it Spitzer} sample, this will give an upper limit on the effect of the missing disk population.
This effect is shown in Figure~\ref{fig:impact}, by the light blue line. It is obvious that this effect is not significant: most of the Unidentified sources seem to be relatively low in mass, and only significantly contribute to the shape of the disk mass distribution below $10\,M_{\oplus}$, where the number of upper limits begins to increase rapidly. This could be due to the southern sub-mosaic being deeper; alternatively, this effect may be a true reflection of the underlying population, and an effect of its relative youth.

\subsubsection{Contamination by free-free contamination at 3\,mm}
\label{sec:free-free}
While other studies generally observes disks in ALMA Band 6 or Band 7 (at $0.8 - 1.3$\,mm), the longer wavelength (3\,mm) at which this sample is observed implies that some of the flux may be contaminated by free-free emission. Indeed, in the ONC, free-free emission from the surface layers of the disk is a significant part of the flux even at higher frequencies~\citep{eisner18}. However, our sample is located at distances greater than 0.5\,pc from the ONC; as a result photoionization of the disk surface is not expected, and our sample in fact does not overlap with the proplyds cataloged in~\citet{ricci08}. Comparison to the 90\,GHz single-dish continuum map from~\citet{dicker09} also shows that in the area surveyed here, free-free emission has dropped off steeply and continuum should dominate. 

However, contamination of our sample by free-free emission from jets cannot be excluded. As a result, we have applied a correction based on the results presented in~\citet{tychoniec18} for the median free-free emission strength and spectral index. At 3\,mm, this is only a small correction of $0.25$\,mJy, and the results of the comparison to other star-forming regions do not change significantly without it.

\subsubsection{The impact of disk temperature and optical depth assumptions}
\label{sec:tempeff}
In this study, we have followed the same approach in converting the flux due to dust to a disk dust mass as previously published work, in order to compare our sample as fairly as possible to those results. However, any of these assumptions may not hold, either generally or in specific environments. Firstly, in the warm environment of Orion, and especially around the ONC, it is not unreasonable to assume that the temperature of the emitting dust even in the well-shielded midplane of the disk is higher than in lower-mass star-forming regions. The ammonia temperature maps of dense pre-stellar gas in Orion presented in~\citet{friesen17}, for instance, show temperatures exceeding 30\,K near the Trapezium. For our data, we assume this is not as important an effect, since our stars are usually located quite far from the ONC (at least 0.5\,pc) and in an environment where cooler gas temperatures (around 20\,K) are measured~\citep{hacar18}. Nonetheless, temperature variations may lead to an overestimate of disk masses in the ONC~\citep{eisner18}, an effect also noted by those authors.

In Figure~\ref{fig:impact}, the mass distribution of disks is shown when assuming a higher temperature of 30\,K. As expected, the effect is to lower the disk masses. However, the size of this effect is only somewhat larger than the scale of the uncertainties in the Kaplan-Meier estimator. In any case, a comparison of the OMC-2 disks with those near the Trapezium should not be affected by this. The Trapezium disks exist in a hotter environment than those in the OMC-2 region and their mass will, therefore, always be overestimated relative to the disk masses of the OMC-2 sample.

An interesting question is if the most massive disks identified in OMC-2 are truly more massive, or simply outliers in effective temperature. The distribution of the disks with dust masses $>100\,M_{\oplus}$ over the field does not seem to favor any particular environment for massive disks. It is possible that these disks have more luminous host stars than other disks in the sample. However, in nearby low-mass star-forming regions, a similar massive tail of the distribution can be present even around fainter stars, like, IM Lup, which is an M0 star~\citep{reipurth96}. At the same time the K$_s$-band data from~\citet{meingast16} for these stars do not suggest them to be particularly luminous, either.

A significant quantity of literature has also been devoted to the question of disk optical depths, with evidence suggesting that a non-negligible fraction of the disk surface is optically thick, suggesting that the disk-integrated flux contains information on both surface area and mass~\citep[e.g.,][]{ricci12,tripathi17}. While this would mean our `dust mass' is incorrect, it would not prevent us from comparing the populations, since we then merely apply some scaling factor to all disk luminosities; it would, however, make the interpretation of what causes a difference between populations more complicated. The recently published results from the high-resolution ALMA Large Program DSHARP have suggested that -- at least for massive disks, which are the objects this survey is sensitive to -- the disk material is optically thin in Band 6 even when trapped in denser rings~\citep{dullemond18} in nearby star-forming regions. If this holds in Orion as well, the assumption of optically thin emission can safely be made in Band 3.

\subsubsection{The effect of contamination by Class I sources}
Throughout this analysis, we have been cautious not to include Class I sources in our Disk sample, in order to make the comparison with other star-forming regions as straightforward as possible. However, the comparison of the~\citet{kainulainen17} catalog shows several protostars misclassified as disks. This suggests the possibility that some contamination is present in our sample, for which we have not yet accounted.

The two most massive disks in this sample are all quite close to groups of cores containing young stellar objects; this might indicate either a misclassification or simply a young age of the disk, leading to a higher mass. Compared to other star-forming regions~\citep[e.g.,][]{ansdell16, pascucci16, barenfeld16}, the most massive disk in our sample is not particularly massive, with a dust mass of $273 \pm 9.8\,M_{\oplus}$, indicating that even these sources are still within the range of masses expected for Class II disks. Removing these sources from the sample has no significant effect on the shape of the disk mass distribution, except for removing the tail of the distribution; we therefore conclude that contamination by younger sources is not likely to be a significant concern for our subsequent analysis.

In conclusion, we have studied the effects of sample incompleteness, free-free contamination, disk temperature assumptions, and contamination with younger, embedded sources on the inferred disk masses. In particular, the completeness of the sample and the free-free subtraction have an opposite effect on the disk mass from the assumed disk temperature. In all cases, the effect on the disk mass distribution is small compared to the errors on the Kaplan-Meier estimator. As such, it is possible to compare this result to other, similar, disk mass distributions.

\section{Discussion}
The disks in OMC-2 are a unique population: on the one hand, they are part of a population of disks around stars that formed in a massive star-forming cloud; on the other, they are at least 0.5\,pc away from the Trapezium, and outside of the region dominated by the radiation field from its massive O-stars. Comparing the disk mass distribution of the OMC-2 population, and that of disks near the Trapezium, therefore allows us to observationally determine a rate of photoevaporative disk mass loss for the first time. Moreover, we can compare this rate to models of external photoevaporation. Simultaneously, comparing this population to that in low-mass star-forming regions allows us to determine if disks evolve in the same way in isolation, or if the cloud from which they originated determines the initial disk mass in some way.

\begin{figure*}[ht]
\begin{center}
		\includegraphics[width=\textwidth]{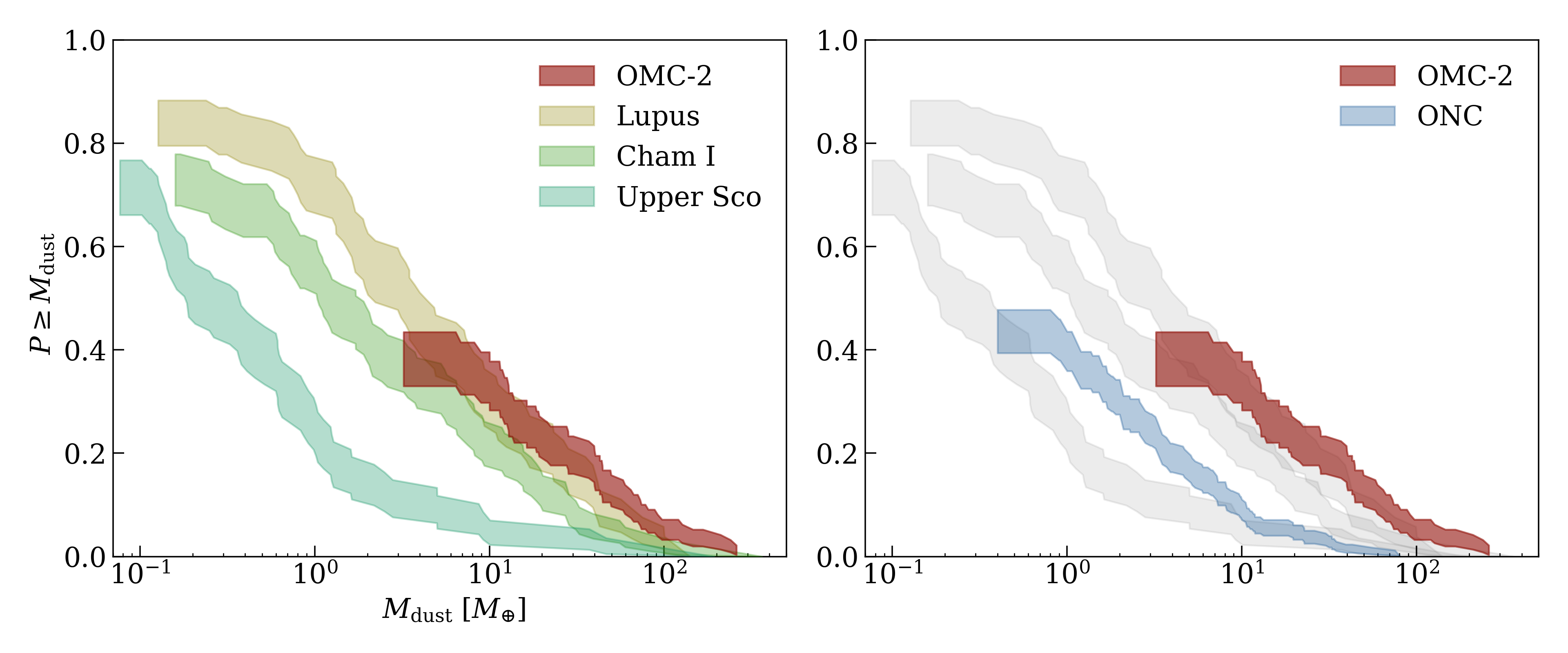}
		\caption{The inferred disk mass distribution in OMC-2 (red) compared to the disk mass distribution in low-mass star forming regions {\it(left} and to disks in the inner $1.5' \times 1.5'$ of the ONC (blue, {\it right}). Data for the low-mass regions comes from~\citet{ansdell16} (Lupus),~\citet{pascucci16} (Cham I), and~\citet{barenfeld16} (Upper Sco); only Class II objects with known stellar masses included. For the ONC, disk masses come from~\citet{eisner18}. The three low-mass regions are reproduced as gray silhouettes in the right panel to facilitate the comparison.}
		\label{fig:massdist}
\end{center}
\end{figure*}

In Figure~\ref{fig:massdist}, the inferred disk mass distribution for our OMC field sample is shown relative to that of several other star-forming regions, derived using a Kaplan-Meier estimator to take the presence of upper limits in the data into account. We do not include the Unidentified source sample for this comparison. In order to compare these populations formally, we use the non-parametric logrank test to compare them without having to make assumptions on the underlying distributions. In addition to the regions shown here, the same procedure was also followed for disk masses in Taurus~\citep{andrews13} and $\sigma$ Ori~\citep{ansdell17}. 

\subsection{Environmental dependence}
The comparison of the OMC-2 population to that in the inner $1.5' \times 1.5'$ of the ONC~\citep{eisner18}, in the right panel of Figure~\ref{fig:massdist}, immediately shows the dramatic effect that the hostile environment of the Trapezium has on disk properties. This is confirmed by a log-rank test: both the ONC sample and the likewise photoevaporated population around $\sigma$ Ori (not shown) are inconsistent with the mass distribution of the OMC-2 disks, with $p \ll 0.05$ in both cases.

While a decrease in disk masses with proximity to the Trapezium has been shown, the OMC-2 sample for the first time allows us to statistically characterize the evolution of disk masses in a dense clustered environment, but outside the main photoevaporation region. To do so, we perform a log-rank test comparing the masses of the ONC population with the masses of the OMC-2 sample multiplied by a constant, given the apparent similarity between the two mass distributions. This leads to a mass ratio of $0.18^{+0.18}_{-0.11}$ between the regions, or a loss of some $80\%$ of the dust mass in 1\,Myr following the age inferred by~\citet{megeath16} for the ONC.

It is interesting to compare the mass loss of these disks to model predictions of the effect of external photoevaporation. In the ONC, and specifically in the inner 0.1\,pc around $\theta^1$ Ori C targeted by~\citet{eisner18}, external photoevaporation by far-UV (FUV) photons is the dominant source of mass loss for disks~\citep{scally01, hollenbach00}, and leads to a mass loss of $\dot{M}_{\text{FUV}} = 2 \times 10^{-9} r_d\,M_{\odot}\,\text{yr}^{-1}$, where $r_d$ is the disk radius in astronomical units. Assuming a typical disk around a solar-mass star in our sample, with a dust mass of 20\,$M_{\oplus}$, an outer radius of 40\,AU, and a gas-to-dust ratio of 100, but which has been exposed to FUV radiation for $0.5$\,Myr, our observations imply a photoevaporative mass loss rate of $1 \times 10^{-8}\,M_{\odot}\,\text{yr}^{-1}$ in the ONC. Importantly, in this approach, the difference in disk temperatures in the ONC and the OMC-2 sample studied here is somewhat degenerate with the age of the region: if the ONC's disks are warmer, the time during which the disk has been exposed to the radiation field can be larger. Comparing this with the result of the equation in~\citet{scally01} the expected mass loss rate for our fiducial disk in the ONC due to FUV radiation is $8 \times 10^{-8}\,M_{\odot}\,\text{yr}^{-1}$. This is a remarkably close correspondence.

To test this result with a more refined model, we have also used the FRIED grid of externally irradiated disks~\citep{haworth18}. Using the same parameters for the disk and a host star mass of $1\,M_{\odot}$ yields a mass loss of $5.25 \times 10^{-8}\,M_{\odot}\,\text{yr}^{-1}$ in a radiation field of $10^4\,G_0$. Here $G_0 = 1.6 \times 10^{-3}$\,erg\,cm$^{-2}$\,s$^{-1}$ is the Habing unit of UV radiation between $912 - 2400$\,\AA, appropriate for the properties of $\theta^1$ Ori C at the distances of the ONC disks~\citep{simondiaz06}. In contrast, the radiation field in the OMC-2 population must be much lower. With purely geometric dilution of the radiation field, based on the average distances of these samples, $G_0$ in the OMC-2 should be a factor $6 \times 10^{-3}$ lower. This is probably an overestimate, given the presence of dust absorption. In this irradiation regime, the FRIED grid predicts very low mass-loss rates, and low sensitivity to variations in the strength of the UV field, for disks with a radius $<100$\,AU. More importantly, the resulting mass loss value is again in good agreement with the calculation made from the observations presented here. This suggests that the primary driver of disk destruction in the ONC is indeed external photoevaporation.

\subsection{Time evolution}
Comparing the OMC-2 disk mass distribution to that of low-mass regions, it becomes obvious that this is a quite massive population (in terms of the most massive quartile), and resembles the young star-forming regions much more closely than the evolved population of disks. Using a log-rank test, we cannot distinguish between the Lupus and OMC-2 disks ($p = 0.87$); the same is true for the comparison to Taurus ($p = 0.76$) and Cham I ($p = 0.2$). However, the tail of the distribution for disk masses in OMC-2 seems to be somewhat more massive than in Taurus and Lupus, and this difference is even bigger when comparing OMC-2 to Cham I. The result of the log-rank tests does not change if we assume a 30\,K effective dust temperature (or, alternatively, a higher opacity). This is also true when comparing the OMC-2 disks to those in Upper Sco: in all cases, the latter population is significantly less massive.

	Including the Unidentified sources that are likely to be disks leads to the OMC field population becoming somewhat more massive, but has no significant effect on the comparison between the disk mass distributions either. Interestingly, it does seem that the shape of these disk mass distributions is now more similar down to our detection limit, even if the average mass increases somewhat. However, our main conclusion is not affected: disks in OMC-2 appear to be most similar to young nearby star-forming regions, and may be somewhat more massive.

The result of this comparison is remarkable given the very different densities, histories, and scales of star formation in these regions as a whole. However, in terms of age, Lupus, Taurus, and the OMC-2 population appear to be quite similar, while Cham I with its slightly lower disk masses is older: the OMC-2 stars are around 1\,Myr in age, compared to Lupus and Taurus at $1-3$\,Myr and $3-5$\,Myr for Cham I. This suggests that, outside of the 0.5\,pc region dominated by the Trapezium, age alone is the dominant factor in protostellar disk evolution.

\subsection{Disk evolution across SFRs}
Our results reveal that studying the OMC-2 disk population is a bridge between disks in the well-studied low-mass star-forming regions and in the dense, highly irradiated environment of the ONC, even with unresolved and relatively shallow observations. It is a population ripe for follow-up analysis at different wavelengths and higher resolution: resolving these disks, in particular, will enable us to compare disk radii as well as masses. In this context, the high-mass tail of disks in the OMC-2 and ONC populations is particularly interesting. Especially in the OMC-2 sample, the inferred masses for these disks are similar to the most massive disks in low-mass star-forming regions, such as TW Hya, IM Lup, and other well-known objects~\citep{andrews16, cleeves16}. These disks are often seen to have dust gaps and rings in nearby SFRs, consistent with planet formation, at scales that can be resolved by ALMA relatively easily even at the distance of Orion A~\citep{andrews18,vdmarel18,vanterwisga18}. Observing such features and studying their relative properties in the OMC-2 and ONC environments would help clarify how environment affects disk substructures. Finally, resolved observations of disks in the OMC-2 cloud will allow us to test more precisely the predictions from theoretical models of disk photoevaporation, which suggest that the radius of a disk is an important factor in the efficiency with which it loses mass.

The result that Class II disk masses, on average, do not sensitively depend on the underlying cloud properties if no strong radiation field is present has important implications for the earlier evolution of circumstellar material. It implies that disk formation can proceed similarly everywhere, with the possible exception of Corona Australis\citep{cazzoletti19} and $\rho$ Ophiuchi~\citep{williams19}. In turn, if planet formation begins early on in the existence of a disk, this too is not strongly affected by environmental factors.

\section{Conclusions}
We have observed 132 protoplanetary disks in the OMC-2 cloud at 3\,mm with ALMA, as well as a population of 35 younger protostars, and use these observations to infer the disk mass distribution in this region.
By placing the masses of disks in OMC-2 in context of both the highly irradiated Trapezium disks, and of nearby low-mass star-forming regions, we can link the evolution of circumstellar material in these very different environments. The massive disks detected in this study are a valuable source of information on disk evolution, and emphasize the power of unbiased surveys, even on relatively shallow datasets. The protostellar sample, while small, is clearly more concentrated along the ISF, tracing local cloud properties, in line with earlier results.

\begin{itemize} 
	\item Dust masses of disks in the OMC field, outside the photoionization region of the Trapezium, reveal a population of protoplanetary disks that appears indistinguishable from the disks in low-mass star forming regions like Lupus and Taurus, which have similar ages but much smaller populations of YSOs than the Orion molecular clouds.
	\item Comparing the OMC-2 sample of disks with the inner $1.5' \times 1.5'$ around the Trapezium reveals a typical mass loss of $\sim 80\%$ of the disk mass in 1\,Myr. The mass ratio between OMC-2 and ONC disks is $0.18^{+0.18}_{-0.11}$; in other words, ONC disks are less massive by a factor $\sim 5$.
	\item Both of these results are robust with respect to our parametrization of disk mass and sample biases.
	\item The difference in mass between the OMC-2 and Trapezium-cluster samples is consistent with disk mass loss driven by far-ultraviolet radiation.
	\item Together, the comparisons to photoevaporated and isolated, low-mass regions imply that in isolation, disk formation and evolution proceeds similarly, regardless of cloud mass.
\end{itemize}

\begin{acknowledgements}
The authors thank the ALLEGRO team for their support during the reduction of the ALMA data. This work is part of the research programme VENI with project number 639.041.644, which is (partly) financed by the Netherlands Organisation for Scientific Research (NWO). AH thanks the Spanish MINECO for support under grant AYA2016- 79006-P. Astrochemistry in Leiden is supported by the European Union A-ERC grant 291141 CHEMPLAN, by the Netherlands Research School for Astronomy (NOVA), and by a Royal Netherlands Academy of Arts and Sciences (KNAW) professor prize. This paper makes use of the following ALMA data: PID 2015.1.00669.S. ALMA is a partnership of ESO (representing its member states) NSF (USA) and NINS (Japan) together with NRC
(Canada) MOST and ASIAA (Taiwan), and KASI (Republic of Korea), in cooperation with the Republic of Chile. The Joint ALMA Observatory is operated by ESO, AUI/NRAO and NAOJ. 
\end{acknowledgements}

\bibliographystyle{aa}
\bibliography{oriondiskbib}

\begin{thebibliography}{49}
\expandafter\ifx\csname natexlab\endcsname\relax\def\natexlab#1{#1}\fi

\bibitem[{{Andrews} {et~al.}(2018){Andrews}, {Huang}, {P{\'e}rez}, {Isella},
  {Dullemond}, {Kurtovic}, {Guzm{\'a}n}, {Carpenter}, {Wilner}, {Zhang}, {Zhu},
  {Birnstiel}, {Bai}, {Benisty}, {Hughes}, {{\"O}berg}, \& {Ricci}}]{andrews18}
{Andrews}, S.~M., {Huang}, J., {P{\'e}rez}, L.~M., {et~al.} 2018, \apj, 869,
  L41

\bibitem[{{Andrews} {et~al.}(2013){Andrews}, {Rosenfeld}, {Kraus}, \&
  {Wilner}}]{andrews13}
{Andrews}, S.~M., {Rosenfeld}, K.~A., {Kraus}, A.~L., \& {Wilner}, D.~J. 2013,
  \apj, 771, 129

\bibitem[{{Andrews} {et~al.}(2016){Andrews}, {Wilner}, {Zhu}, {Birnstiel},
  {Carpenter}, {P{\'e}rez}, {Bai}, {{\"O}berg}, {Hughes}, {Isella}, \&
  {Ricci}}]{andrews16}
{Andrews}, S.~M., {Wilner}, D.~J., {Zhu}, Z., {et~al.} 2016, \apj, 820, L40

\bibitem[{{Ansdell} {et~al.}(2017){Ansdell}, {Williams}, {Manara}, {Miotello},
  {Facchini}, {van der Marel}, {Testi}, \& {van Dishoeck}}]{ansdell17}
{Ansdell}, M., {Williams}, J.~P., {Manara}, C.~F., {et~al.} 2017, \aj, 153, 240

\bibitem[{{Ansdell} {et~al.}(2016){Ansdell}, {Williams}, {van der Marel},
  {Carpenter}, {Guidi}, {Hogerheijde}, {Mathews}, {Manara}, {Miotello},
  {Natta}, {Oliveira}, {Tazzari}, {Testi}, {van Dishoeck}, \& {van
  Terwisga}}]{ansdell16}
{Ansdell}, M., {Williams}, J.~P., {van der Marel}, N., {et~al.} 2016, \apj,
  828, 46

\bibitem[{{Barenfeld} {et~al.}(2016){Barenfeld}, {Carpenter}, {Ricci}, \&
  {Isella}}]{barenfeld16}
{Barenfeld}, S.~A., {Carpenter}, J.~M., {Ricci}, L., \& {Isella}, A. 2016,
  \apj, 827, 142

\bibitem[{{Baruteau} {et~al.}(2014){Baruteau}, {Crida}, {Paardekooper},
  {Masset}, {Guilet}, {Bitsch}, {Nelson}, {Kley}, \& {Papaloizou}}]{baruteau14}
{Baruteau}, C., {Crida}, A., {Paardekooper}, S.~J., {et~al.} 2014, in
  Protostars and Planets VI, ed. H.~{Beuther}, R.~S. {Klessen}, C.~P.
  {Dullemond}, \& T.~{Henning}, 667

\bibitem[{{Beckwith} {et~al.}(1990){Beckwith}, {Sargent}, {Chini}, \&
  {Guesten}}]{beckwith90}
{Beckwith}, S. V.~W., {Sargent}, A.~I., {Chini}, R.~S., \& {Guesten}, R. 1990,
  \aj, 99, 924

\bibitem[{{Benz} {et~al.}(2014){Benz}, {Ida}, {Alibert}, {Lin}, \&
  {Mordasini}}]{benz14}
{Benz}, W., {Ida}, S., {Alibert}, Y., {Lin}, D., \& {Mordasini}, C. 2014, in
  Protostars and Planets VI, ed. H.~{Beuther}, R.~S. {Klessen}, C.~P.
  {Dullemond}, \& T.~{Henning}, 691

\bibitem[{{Carpenter}(2000)}]{carpenter00}
{Carpenter}, J.~M. 2000, \aj, 120, 3139

\bibitem[{{Cazzoletti} {et~al.}(2019){Cazzoletti}, {Manara}, {Liu}, {van
  Dishoeck}, {Facchini}, {Alcal{\`a}}, {Ansdell}, {Testi}, {Williams},
  {Carrasco-Gonz{\'a}lez}, {Dong}, {Forbrich}, {Fukagawa}, {Galv{\'a}n-Madrid},
  {Hirano}, {Hogerheijde}, {Hasegawa}, {Muto}, {Pinilla}, {Takami}, {Tamura},
  {Tazzari}, \& {Wisniewski}}]{cazzoletti19}
{Cazzoletti}, P., {Manara}, C.~F., {Liu}, H.~B., {et~al.} 2019, arXiv e-prints,
  arXiv:1904.02409

\bibitem[{{Cleeves} {et~al.}(2016){Cleeves}, {{\"O}berg}, {Wilner}, {Huang},
  {Loomis}, {Andrews}, \& {Czekala}}]{cleeves16}
{Cleeves}, L.~I., {{\"O}berg}, K.~I., {Wilner}, D.~J., {et~al.} 2016, \apj,
  832, 110

\bibitem[{{Dicker} {et~al.}(2009){Dicker}, {Mason}, {Korngut}, {Cotton},
  {Compi{\`e}gne}, {Devlin}, {Martin}, {Ade}, {Benford}, {Irwin}, {Maddalena},
  {McMullin}, {Shepherd}, {Sievers}, {Staguhn}, \& {Tucker}}]{dicker09}
{Dicker}, S.~R., {Mason}, B.~S., {Korngut}, P.~M., {et~al.} 2009, \apj, 705,
  226

\bibitem[{{Dullemond} {et~al.}(2018){Dullemond}, {Birnstiel}, {Huang},
  {Kurtovic}, {Andrews}, {Guzm{\'a}n}, {P{\'e}rez}, {Isella}, {Zhu}, {Benisty},
  {Wilner}, {Bai}, {Carpenter}, {Zhang}, \& {Ricci}}]{dullemond18}
{Dullemond}, C.~P., {Birnstiel}, T., {Huang}, J., {et~al.} 2018, arXiv
  e-prints, arXiv:1812.04044

\bibitem[{{Eisner} {et~al.}(2018){Eisner}, {Arce}, {Ballering}, {Bally},
  {Andrews}, {Boyden}, {Di Francesco}, {Fang}, {Johnstone}, {Kim}, {Mann},
  {Matthews}, {Pascucci}, {Ricci}, {Sheehan}, \& {Williams}}]{eisner18}
{Eisner}, J.~A., {Arce}, H.~G., {Ballering}, N.~P., {et~al.} 2018, \apj, 860,
  77

\bibitem[{{Eisner} {et~al.}(2008){Eisner}, {Plambeck}, {Carpenter}, {Corder},
  {Qi}, \& {Wilner}}]{eisner08}
{Eisner}, J.~A., {Plambeck}, R.~L., {Carpenter}, J.~M., {et~al.} 2008, \apj,
  683, 304

\bibitem[{{Evans} {et~al.}(2003){Evans}, {Allen}, {Blake}, {Boogert}, {Bourke},
  {Harvey}, {Kessler}, {Koerner}, {Lee}, {Mundy}, {Myers}, {Padgett},
  {Pontoppidan}, {Sargent}, {Stapelfeldt}, {van Dishoeck}, {Young}, \&
  {Young}}]{evans03}
{Evans}, II, N.~J., {Allen}, L.~E., {Blake}, G.~A., {et~al.} 2003, \pasp, 115,
  965

\bibitem[{{Felli} {et~al.}(1993){Felli}, {Churchwell}, {Wilson}, \&
  {Taylor}}]{felli93}
{Felli}, M., {Churchwell}, E., {Wilson}, T.~L., \& {Taylor}, G.~B. 1993, \aaps,
  98, 137

\bibitem[{{Friesen} {et~al.}(2017){Friesen}, {Pineda}, {co-PIs}, {Rosolowsky},
  {Alves}, {Chac{\'o}n-Tanarro}, {How-Huan Chen}, {Chun-Yuan Chen}, {Di
  Francesco}, {Keown}, {Kirk}, {Punanova}, {Seo}, {Shirley}, {Ginsburg},
  {Hall}, {Offner}, {Singh}, {Arce}, {Caselli}, {Goodman}, {Martin}, {Matzner},
  {Myers}, {Redaelli}, \& {GAS Collaboration}}]{friesen17}
{Friesen}, R.~K., {Pineda}, J.~E., {co-PIs}, {et~al.} 2017, \apj, 843, 63

\bibitem[{{Furlan} {et~al.}(2016){Furlan}, {Fischer}, {Ali}, {Stutz}, {Stanke},
  {Tobin}, {Megeath}, {Osorio}, {Hartmann}, {Calvet}, {Poteet}, {Booker},
  {Manoj}, {Watson}, \& {Allen}}]{furlan16}
{Furlan}, E., {Fischer}, W.~J., {Ali}, B., {et~al.} 2016, The Astrophysical
  Journal Supplement Series, 224, 5

\bibitem[{{Getman} {et~al.}(2005){Getman}, {Flaccomio}, {Broos}, {Grosso},
  {Tsujimoto}, {Townsley}, {Garmire}, {Kastner}, {Li}, {Harnden}, {Wolk},
  {Murray}, {Lada}, {Muench}, {McCaughrean}, {Meeus}, {Damiani}, {Micela},
  {Sciortino}, {Bally}, {Hillenbrand }, {Herbst}, {Preibisch}, \&
  {Feigelson}}]{getman05}
{Getman}, K.~V., {Flaccomio}, E., {Broos}, P.~S., {et~al.} 2005, \apjs, 160,
  319

\bibitem[{{Gro{\ss}schedl} {et~al.}(2019){Gro{\ss}schedl}, {Alves}, {Teixeira},
  {Bouy}, {Forbrich}, {Lada}, {Meingast}, {Hacar}, {Ascenso}, {Ackerl},
  {Hasenberger}, {K{\"o}hler}, {Kubiak}, {Larreina}, {Linhardt}, {Lombardi}, \&
  {M{\"o}ller}}]{grossschedl18}
{Gro{\ss}schedl}, J.~E., {Alves}, J., {Teixeira}, P.~S., {et~al.} 2019, \aap,
  622, A149

\bibitem[{{Hacar} {et~al.}(2018){Hacar}, {Tafalla}, {Forbrich}, {Alves},
  {Meingast}, {Grossschedl}, \& {Teixeira}}]{hacar18}
{Hacar}, A., {Tafalla}, M., {Forbrich}, J., {et~al.} 2018, \aap, 610, A77

\bibitem[{{Haworth} {et~al.}(2018){Haworth}, {Clarke}, {Rahman}, {Winter}, \&
  {Facchini}}]{haworth18}
{Haworth}, T.~J., {Clarke}, C.~J., {Rahman}, W., {Winter}, A.~J., \&
  {Facchini}, S. 2018, \mnras, 481, 452

\bibitem[{{Hollenbach} {et~al.}(2000){Hollenbach}, {Yorke}, \&
  {Johnstone}}]{hollenbach00}
{Hollenbach}, D.~J., {Yorke}, H.~W., \& {Johnstone}, D. 2000, in Protostars and
  Planets IV, ed. V.~{Mannings}, A.~P. {Boss}, \& S.~S. {Russell}, 401--428

\bibitem[{{Kainulainen} {et~al.}(2017){Kainulainen}, {Stutz}, {Stanke}, {Abreu-
  Vicente}, {Beuther}, {Henning}, {Johnston}, \& {Megeath}}]{kainulainen17}
{Kainulainen}, J., {Stutz}, A.~M., {Stanke}, T., {et~al.} 2017, \aap, 600, A141

\bibitem[{{Lada} {et~al.}(1993){Lada}, {Strom}, \& {Myers}}]{lada93}
{Lada}, E.~A., {Strom}, K.~M., \& {Myers}, P.~C. 1993, in Protostars and
  Planets III, ed. E.~H. {Levy} \& J.~I. {Lunine}, 245

\bibitem[{{Mairs} {et~al.}(2019){Mairs}, {Lalchand}, {Bower}, {Forbrich},
  {Bell}, {Herczeg}, {Johnstone}, {Chen}, {Lee}, \& {Hacar}}]{mairs19}
{Mairs}, S., {Lalchand}, B., {Bower}, G.~C., {et~al.} 2019, \apj, 871, 72

\bibitem[{{Mann} {et~al.}(2014){Mann}, {Di Francesco}, {Johnstone}, {Andrews},
  {Williams}, {Bally}, {Ricci}, {Hughes}, \& {Matthews}}]{mann14}
{Mann}, R.~K., {Di Francesco}, J., {Johnstone}, D., {et~al.} 2014, \apj, 784,
  82

\bibitem[{{Mann} \& {Williams}(2010)}]{mann10}
{Mann}, R.~K. \& {Williams}, J.~P. 2010, \apj, 725, 430

\bibitem[{{Megeath} {et~al.}(2012){Megeath}, {Gutermuth}, {Muzerolle},
  {Kryukova}, {Flaherty}, {Hora}, {Allen}, {Hartmann}, {Myers}, {Pipher},
  {Stauffer}, {Young}, \& {Fazio}}]{megeath12}
{Megeath}, S.~T., {Gutermuth}, R., {Muzerolle}, J., {et~al.} 2012, \aj, 144,
  192

\bibitem[{{Megeath} {et~al.}(2016){Megeath}, {Gutermuth}, {Muzerolle},
  {Kryukova}, {Hora}, {Allen}, {Flaherty}, {Hartmann}, {Myers}, {Pipher},
  {Stauffer}, {Young}, \& {Fazio}}]{megeath16}
{Megeath}, S.~T., {Gutermuth}, R., {Muzerolle}, J., {et~al.} 2016, \aj, 151, 5

\bibitem[{{Meingast} {et~al.}(2016){Meingast}, {Alves}, {Mardones}, {Teixeira},
  {Lombardi}, {Gro{\ss}schedl}, {Ascenso}, {Bouy}, {Forbrich}, {Goodman},
  {Hacar}, {Hasenberger}, {Kainulainen}, {Kubiak}, {Lada}, {Lada}, {Moitinho},
  {Petr-Gotzens}, {Rodrigues}, \& {Rom{\'a}n-Z{\'u}{\~n}iga}}]{meingast16}
{Meingast}, S., {Alves}, J., {Mardones}, D., {et~al.} 2016, \aap, 587, A153

\bibitem[{{Menten} {et~al.}(2007){Menten}, {Reid}, {Forbrich}, \&
  {Brunthaler}}]{menten07}
{Menten}, K.~M., {Reid}, M.~J., {Forbrich}, J., \& {Brunthaler}, A. 2007, \aap,
  474, 515

\bibitem[{{O'Dell}(1998)}]{odell98}
{O'Dell}, C.~R. 1998, \aj, 115, 263

\bibitem[{{O'Dell} {et~al.}(1993){O'Dell}, {Wen}, \& {Hu}}]{odell93}
{O'Dell}, C.~R., {Wen}, Z., \& {Hu}, X. 1993, \apj, 410, 696

\bibitem[{{Pascucci} {et~al.}(2016){Pascucci}, {Testi}, {Herczeg}, {Long},
  {Manara}, {Hendler}, {Mulders}, {Krijt}, {Ciesla}, {Henning}, {Mohanty},
  {Drabek-Maunder}, {Apai}, {Sz{\'{u}}cs}, {Sacco}, \& {Olofsson}}]{pascucci16}
{Pascucci}, I., {Testi}, L., {Herczeg}, G.~J., {et~al.} 2016, \apj, 831, 125

\bibitem[{{Peterson} \& {Megeath}(2008)}]{peterson08}
{Peterson}, D.~E. \& {Megeath}, S.~T. 2008, {The Orion Molecular Cloud 2/3 and
  NGC 1977 Regions} (Astronomical Society of the Pacific Monograph
  Publications), 590

\bibitem[{{Reipurth} {et~al.}(1996){Reipurth}, {Pedrosa}, \&
  {Lago}}]{reipurth96}
{Reipurth}, B., {Pedrosa}, A., \& {Lago}, M.~T.~V.~T. 1996, \aaps, 120, 229

\bibitem[{{Ricci} {et~al.}(2008){Ricci}, {Robberto}, \& {Soderblom}}]{ricci08}
{Ricci}, L., {Robberto}, M., \& {Soderblom}, D.~R. 2008, \aj, 136, 2136

\bibitem[{{Ricci} {et~al.}(2012){Ricci}, {Trotta}, {Testi}, {Natta}, {Isella},
  \& {Wilner}}]{ricci12}
{Ricci}, L., {Trotta}, F., {Testi}, L., {et~al.} 2012, \aap, 540, A6

\bibitem[{{Scally} \& {Clarke}(2001)}]{scally01}
{Scally}, A. \& {Clarke}, C. 2001, \mnras, 325, 449

\bibitem[{{Sim{\'o}n-D{\'\i}az} {et~al.}(2006){Sim{\'o}n-D{\'\i}az}, {Herrero},
  {Esteban}, \& {Najarro}}]{simondiaz06}
{Sim{\'o}n-D{\'\i}az}, S., {Herrero}, A., {Esteban}, C., \& {Najarro}, F. 2006,
  \aap, 448, 351

\bibitem[{{Tripathi} {et~al.}(2017){Tripathi}, {Andrews}, {Birnstiel}, \&
  {Wilner}}]{tripathi17}
{Tripathi}, A., {Andrews}, S.~M., {Birnstiel}, T., \& {Wilner}, D.~J. 2017,
  \apj, 845, 44

\bibitem[{{Tychoniec} {et~al.}(2018){Tychoniec}, {Tobin}, {Karska}, {Chandler},
  {Dunham}, {Harris}, {Kratter}, {Li}, {Looney}, {Melis}, {P{\'e}rez},
  {Sadavoy}, {Segura-Cox}, \& {van Dishoeck}}]{tychoniec18}
{Tychoniec}, {\L}., {Tobin}, J.~J., {Karska}, A., {et~al.} 2018, The
  Astrophysical Journal Supplement Series, 238, 19

\bibitem[{{van der Marel} {et~al.}(2018){van der Marel}, {Williams}, \&
  {Bruderer}}]{vdmarel18}
{van der Marel}, N., {Williams}, J.~P., \& {Bruderer}, S. 2018, \apj, 867, L14

\bibitem[{{van Terwisga} {et~al.}(2018){van Terwisga}, {van Dishoeck},
  {Ansdell}, {van der Marel}, {Testi}, {Williams}, {Facchini}, {Tazzari},
  {Hogerheijde}, {Trapman}, {Manara}, {Miotello}, {Maud}, \&
  {Harsono}}]{vanterwisga18}
{van Terwisga}, S.~E., {van Dishoeck}, E.~F., {Ansdell}, M., {et~al.} 2018,
  \aap, 616, A88

\bibitem[{{Williams} {et~al.}(2019){Williams}, {Cieza}, {Hales}, {Ansdell},
  {Ruiz-Rodriguez}, {Casassus}, {Perez}, \& {Zurlo}}]{williams19}
{Williams}, J.~P., {Cieza}, L., {Hales}, A., {et~al.} 2019, \apj, 875, L9

\bibitem[{{Williams} \& {Gaidos}(2007)}]{williams07}
{Williams}, J.~P. \& {Gaidos}, E. 2007, \apj, 663, L33

\end{thebibliography}

\begin{appendix}
\section{Fluxes of additional sources identified in the ALMA data}
\setcounter{table}{0}
\label{app:unidentified}

In the following Tables, the fluxes, masses, and positions for the protostellar sources (Table~\ref{tab:P}) are listed, as well as the fluxes and positions for the sample of unknown sources that are detected with ALMA but not in the Spitzer catalog (Table~\ref{tab:U}). Please note that the masses for the detected protostars are calculated with the same assumptions that were made for the disks, but that these objects are expected to have significant contamination from envelope material. Finally, for the protostellar sources, some flux may be resolved-out.

\begin{table*}
\caption{Continuum fluxes and masses for the detected protostars}
\label{tab:P}
\centering
\begin{tabular}{l l l l l}
\hline \hline
ID & RA & Dec & Flux & Mass \\
 & J2000 & J2000 & mJy & $M_{\oplus}$ \\
\hline
P1 & 05:35:27.63 &-05:09:33.5 & $23.1 \pm 0.16$ & $1316.6 \pm 9.45$ \\
P2 & 05:35:20.14 &-05:13:15.5 & $21.6 \pm 0.17$ & $1229.0 \pm 9.84$ \\
P3 & 05:35:23.33 &-05:12:03.1 & $17.8 \pm 0.16$ & $1010.3 \pm 9.03$ \\
P4 & 05:35:25.82 &-05:05:43.6 & $15.0 \pm 0.24$ & $850.9 \pm 14.07$ \\
P5 & 05:35:19.84 &-05:15:08.5 & $9.2 \pm 0.12$ & $516.8 \pm 6.69$ \\
P6 & 05:35:24.30 &-05:08:30.6 & $9.0 \pm 0.17$ & $502.2 \pm 9.88$ \\
P7 & 05:35:25.61 &-05:07:57.3 & $8.6 \pm 0.17$ & $478.9 \pm 9.92$ \\
P8 & 05:35:25.23 &-05:15:35.7 & $8.4 \pm 0.13$ & $467.4 \pm 7.22$ \\
P9 & 05:35:24.73 &-05:10:30.2 & $7.5 \pm 0.16$ & $418.4 \pm 9.20$ \\
P10 & 05:35:26.84 &-05:09:24.6 & $2.9 \pm 0.17$ & $150.6 \pm 9.73$ \\
P11 & 05:35:18.20 &-05:13:06.7 & $2.8 \pm 0.18$ & $148.4 \pm 10.24$ \\
P12 & 05:35:27.00 &-05:09:54.1 & $2.2 \pm 0.16$ & $110.5 \pm 9.41$ \\
P13 & 05:35:21.55 &-05:09:38.7 & $2.1 \pm 0.18$ & $104.7 \pm 10.42$ \\
P14 & 05:35:20.73 &-05:13:23.6 & $2.1 \pm 0.17$ & $104.2 \pm 9.60$ \\
P15 & 05:35:27.07 &-05:10:00.4 & $2.0 \pm 0.16$ & $101.8 \pm 9.35$ \\
P16 & 05:35:19.47 &-05:15:32.7 & $1.8 \pm 0.11$ & $89.3 \pm 6.15$ \\
P17 & 05:35:24.86 &-05:06:21.4 & $1.7 \pm 0.18$ & $82.6 \pm 10.18$ \\
P18 & 05:35:21.87 &-05:07:01.6 & $1.6 \pm 0.19$ & $80.0 \pm 10.72$ \\
P19 & 05:35:27.70 &-05:07:03.5 & $1.3 \pm 0.17$ & $57.9 \pm 10.07$ \\
P20 & 05:35:24.58 &-05:11:29.7 & $1.2 \pm 0.15$ & $54.0 \pm 8.61$ \\
P21 & 05:35:26.66 &-05:06:10.3 & $1.2 \pm 0.19$ & $53.4 \pm 10.70$ \\
P22 & 05:35:26.97 &-05:10:17.2 & $0.7 \pm 0.16$ & $27.5 \pm 9.20$ \\
P23 & 05:35:12.02 &-05:18:40.8 & $0.7 \pm 0.12$ & $23.9 \pm 6.82$ \\
P24 & 05:35:26.20 &-05:08:33.4 & $0.6 \pm 0.17$ & $22.8 \pm 10.04$ \\
P25 & 05:35:11.61 &-05:19:12.4 & $0.4 \pm 0.12$ & $11.2 \pm 6.87$ \\
P26 & 05:35:22.10 &-05:18:57.7 & $0.3 \pm 0.11$ & $5.5 \pm 6.50$ \\
\hline
\end{tabular}
\end{table*}

\begin{table*}
\caption{Continuum fluxes and masses for objects not included in the {\it Spitzer} catalog}
\label{tab:U}
\centering
\begin{tabular}{l l l l l}
\hline \hline
ID & RA & Dec & Flux & Mass \\
 &  &  & mJy & $M_{\oplus}$ \\
\hline
U1\tablefootmark{a} & 05:35:21.36 &-05:13:17.8 & $10.5 \pm 0.16$ & $590.1 \pm 9.37$ \\
U2\tablefootmark{a} & 05:35:26.42 &-05:10:06.1 & $6.2 \pm 0.16$ & $344.1 \pm 9.34$ \\
U3\tablefootmark{a} & 05:35:24.91 &-05:07:54.3 & $5.9 \pm 0.17$ & $323.7 \pm 9.84$ \\
U4\tablefootmark{a} & 05:35:19.43 &-05:15:38.3 & $4.6 \pm 0.11$ & $249.1 \pm 6.16$ \\
U5\tablefootmark{a} & 05:35:22.77 &-05:12:27.0 & $3.2 \pm 0.16$ & $170.9 \pm 9.30$ \\
U6\tablefootmark{b,}\tablefootmark{c} & 05:35:22.74 &-05:18:38.2 & $2.4 \pm 0.12$ & $122.9 \pm 6.65$ \\
U7\tablefootmark{a} & 05:35:27.02 &-05:09:49.0 & $1.3 \pm 0.16$ & $63.1 \pm 9.46$ \\
U8\tablefootmark{a} & 05:35:26.81 &-05:10:05.7 & $1.2 \pm 0.16$ & $52.0 \pm 9.32$ \\
U9\tablefootmark{c} & 05:35:21.72 &-05:17:40.2 & $1.1 \pm 0.11$ & $46.5 \pm 6.40$ \\
U10\tablefootmark{a} & 05:35:25.27 &-05:07:49.8 & $0.8 \pm 0.17$ & $33.3 \pm 9.92$ \\
U11\tablefootmark{c} & 05:35:13.25 &-05:18:32.8 & $0.8 \pm 0.11$ & $32.8 \pm 6.61$ \\
U12 & 05:35:23.07 &-05:11:50.6 & $0.8 \pm 0.16$ & $32.8 \pm 9.17$ \\
U13\tablefootmark{b,}\tablefootmark{c} & 05:35:23.40 &-05:18:50.8 & $0.7 \pm 0.14$ & $27.8 \pm 7.87$ \\
U14 & 05:35:20.48 &-05:18:15.3 & $0.6 \pm 0.11$ & $22.2 \pm 6.61$ \\
U15\tablefootmark{a} & 05:35:21.72 &-05:13:12.9 & $0.6 \pm 0.16$ & $20.6 \pm 9.22$ \\
U16\tablefootmark{b} & 05:35:19.85 &-05:18:41.0 & $0.6 \pm 0.11$ & $19.1 \pm 6.54$ \\
U17 & 05:35:26.27 &-05:09:55.4 & $0.6 \pm 0.16$ & $19.0 \pm 9.44$ \\
U18 & 05:35:28.37 &-05:09:26.1 & $0.6 \pm 0.16$ & $18.2 \pm 9.20$ \\
U19\tablefootmark{a} & 05:35:22.98 &-05:12:40.0 & $0.6 \pm 0.16$ & $17.8 \pm 9.05$ \\
U20 & 05:35:26.39 &-05:10:21.9 & $0.5 \pm 0.16$ & $16.4 \pm 9.18$ \\
U21\tablefootmark{b,}\tablefootmark{c} & 05:35:17.44 &-05:17:40.6 & $0.5 \pm 0.11$ & $15.7 \pm 6.40$ \\
U22 & 05:35:20.84 &-05:13:16.1 & $0.5 \pm 0.17$ & $15.6 \pm 9.61$ \\
U23\tablefootmark{c} & 05:35:16.99 &-05:17:32.1 & $0.5 \pm 0.11$ & $14.4 \pm 6.34$ \\
U24 & 05:35:21.11 &-05:17:15.5 & $0.5 \pm 0.11$ & $12.3 \pm 6.43$ \\
U25\tablefootmark{c} & 05:35:14.43 &-05:18:49.0 & $0.4 \pm 0.11$ & $10.9 \pm 6.39$ \\
U26\tablefootmark{c} & 05:35:14.73 &-05:16:52.0 & $0.4 \pm 0.11$ & $9.7 \pm 6.41$ \\
U27 & 05:35:14.43 &-05:18:58.5 & $0.4 \pm 0.11$ & $9.6 \pm 6.39$ \\
U28\tablefootmark{b,}\tablefootmark{c} & 05:35:20.00 &-05:18:46.4 & $0.4 \pm 0.11$ & $8.4 \pm 6.52$ \\
U29 & 05:35:14.64 &-05:18:37.3 & $0.4 \pm 0.11$ & $8.2 \pm 6.36$ \\
U30\tablefootmark{c} & 05:35:24.40 &-05:14:58.2 & $0.4 \pm 0.13$ & $8.2 \pm 7.26$ \\
U31 & 05:35:21.93 &-05:15:44.5 & $0.4 \pm 0.10$ & $7.8 \pm 5.84$ \\
U32 & 05:35:17.83 &-05:18:49.5 & $0.4 \pm 0.11$ & $7.2 \pm 6.38$ \\
U33\tablefootmark{b,}\tablefootmark{c} & 05:35:21.90 &-05:17:03.3 & $0.4 \pm 0.11$ & $6.8 \pm 6.52$ \\
U34\tablefootmark{b,}\tablefootmark{c} & 05:35:14.55 &-05:17:25.3 & $0.4 \pm 0.11$ & $6.0 \pm 6.48$ \\
\hline
\end{tabular}
\tablefoot{
\tablefoottext{a}{Source detected in~\citet{kainulainen17}}
\tablefoottext{b}{Source detected in COUPS~\citep{getman05}}
\tablefoottext{c}{Source detected in VISION~\citep{meingast16}}
}
\end{table*}

\renewcommand\thefigure{\thesection.\arabic{figure}}
\section{Cutouts of detected sources}
\setcounter{figure}{0}
\label{app:cutouts}

\begin{figure*}[ht]
\centering
		\includegraphics[width=17cm]{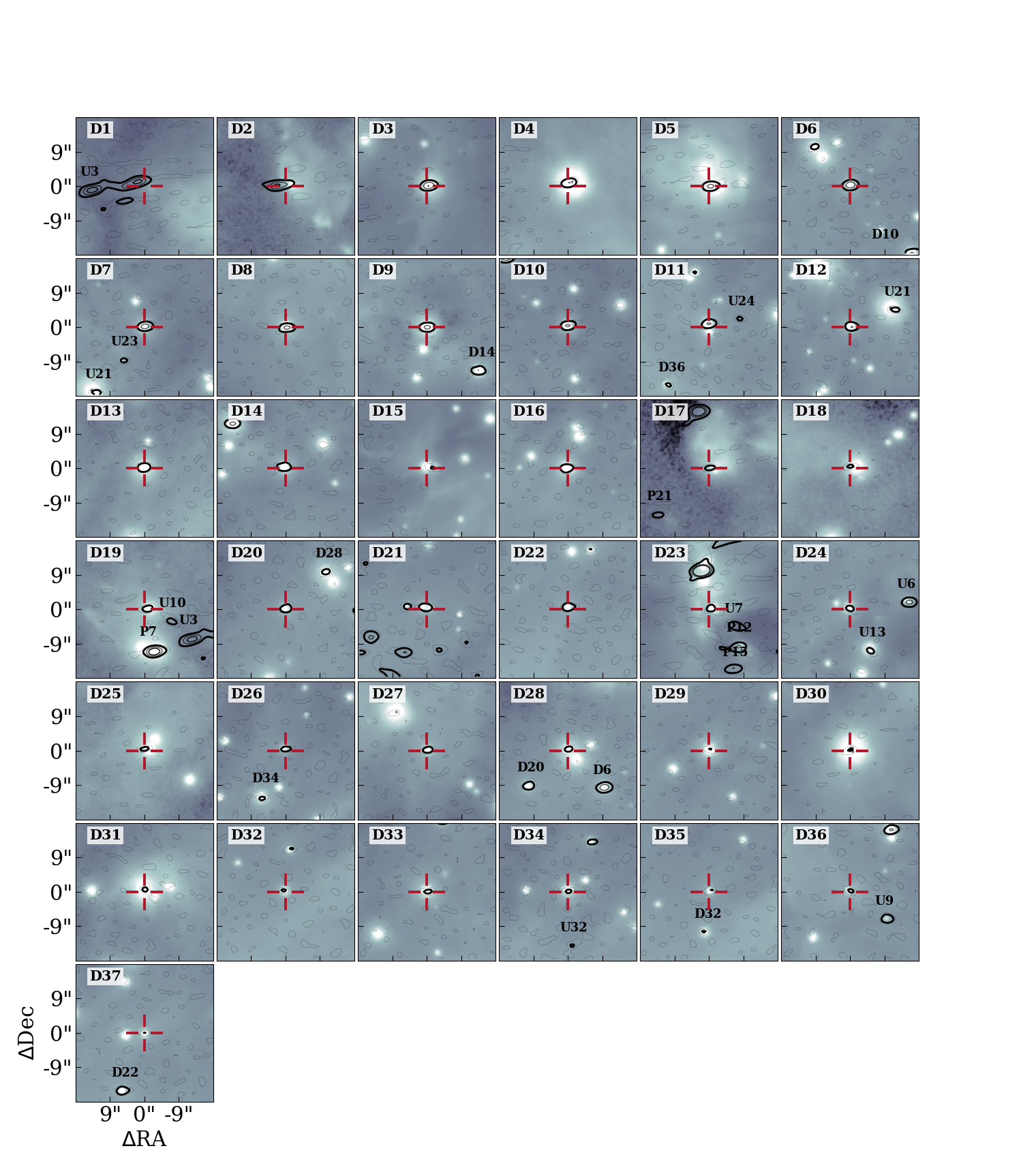}
		\caption{All disks detected in {\it Spitzer}~\citep{megeath12} and ALMA (this study), ordered by millimeter flux. Solid contours show the 3, 12 and 24$\sigma$ noise levels of the ALMA observations; $1\sigma$ flux contours are dashed. the background is the VISTA K$_s$ image from~\citet{meingast16}. Small labels indicate the types of nearby objects in this survey.}
		\label{fig:diskstamps}
\end{figure*}

\begin{figure*}[ht]
\centering
		\includegraphics[width=17cm]{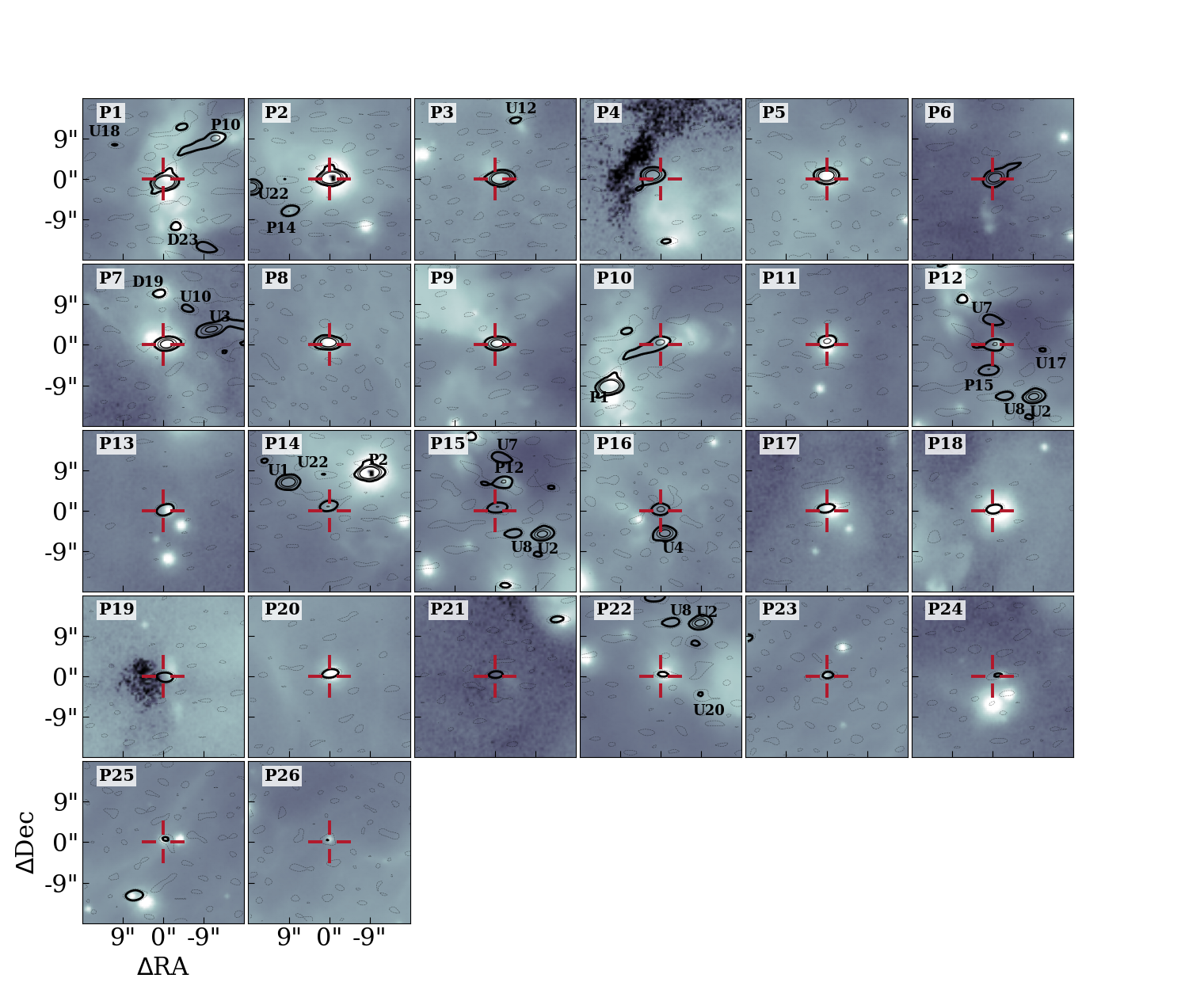}
		\caption{All protostars detected in {\it Spitzer}~\citep{megeath12} and ALMA (this study), ordered by millimeter flux. Solid contours show the 3, 12 and 24$\sigma$ noise levels of the ALMA observations; $1\sigma$ flux contours are dashed. the background is the VISTA K$_s$ image from~\citet{meingast16}. Small labels indicate the types of nearby objects in this survey.}
		\label{fig:protstamps}
\end{figure*}

\begin{figure*}[ht]
\centering
		\includegraphics[width=17cm]{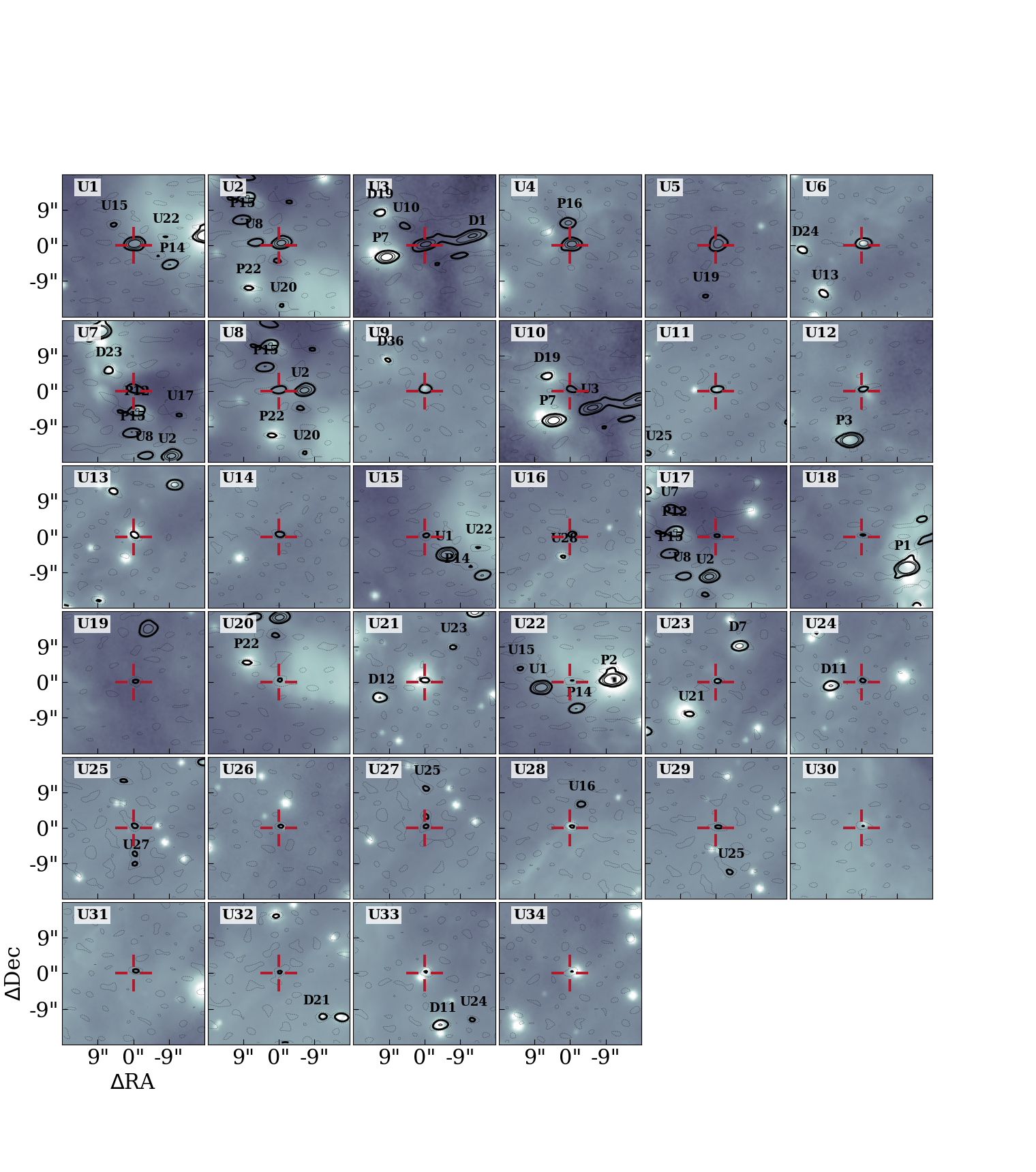}
		\caption{All point sources detected by ALMA (this study) but missing {\it Spitzer} counterparts~\citep{megeath12}, ordered by millimeter flux. Solid contours show the 3, 6, 12 and 24$\sigma$ noise levels of the ALMA observations; $1\sigma$ flux contours are dashed. the background is the VISTA K$_s$ image from~\citet{meingast16}. Small labels indicate the types of nearby objects in this survey.}
		\label{fig:ustamps}
\end{figure*}

\end{appendix}

\end{document}